\documentclass{aastex62}
\usepackage{subfigure}
\graphicspath{{./}{figures/}}

\received{January 1, 2019}
\revised{January 7, 2019}
\accepted{\today}
\shorttitle{Enhanced neutrino emissivities in pseudoscalar-mediated Dark Matter annihilation in neutron stars}
\shortauthors{Cerme\~no et al.}
 \usepackage[bookmarks=false]{hyperref}

\begin{document}

\title{Enhanced neutrino emissivities in pseudoscalar-mediated \\ Dark Matter annihilation in neutron stars}

\author{M. Cerme\~no$^1$~\footnote{marinacgavilan@usal.es}, M. A. P\'erez-Garc\'ia$^1$~\footnote{mperezga@usal.es} and R. A.  Lineros$^{2}$~\footnote{rlineros@gmail.com}}

\affiliation{$^1$ Department of Fundamental Physics, University of Salamanca, Plaza de la Merced s/n, E-37008 Salamanca, Spain\\ $^2$ Instituto de F\'{\i}sica Corpuscular -- CSIC/U. Valencia,\\ Parc Cient\'{\i}fic, calle Catedr\'{a}tico Jos\'{e} Beltr\'{a}n 2, E-46980 Paterna, Spain.\\}

\begin{abstract}
We calculate neutrino emissivities from self-annihilating dark matter ($\chi$) in the dense and hot stellar interior of a (proto)neutron star. Using a model where dark matter interacts with nucleons in the stellar core through a pseudoscalar boson ($a$) we find that the neutrino production rates from the dominant reaction channels $\chi \chi \rightarrow \nu \bar{\nu}$ or $\chi \chi \rightarrow a a$, with subsequent decay of the mediator $ a \rightarrow \nu \bar{\nu}$, could locally match and even surpass those of the standard neutrinos from the modified nuclear URCA processes at early ages. We find that the emitting region can be localized in a tiny fraction of the star (less than a few percent of the core volume) and the process can last its entire lifetime for some cases under study. We discuss the possible consequences of our results for stellar cooling in light of existing dark matter constraints.
\end{abstract}

\begin{keywords}
{dark matter, neutron star, neutrinos, emissivity}
\end{keywords}

%%%%%%%%%%%%%
%%%%%%%%%%%%%
%%% SECTION 1 %%%%
%%%%%%%%%%%%%
%%%%%%%%%%%%%

\section{Introduction}
\bigskip
Dark Matter (DM) is an essential ingredient of the standard cosmological model. We now know it constitutes nearly 85\% of the Universe matter density. However, despite the tremendous amount of progress that has been made in the search for this missing type of matter, both on the theoretical and experimental fronts, its true nature remains an open question. The Standard Model (SM) of particle physics alone cannot explain the nature of DM, suggesting that it must be extended. Many theoretical model proposals have arisen aiming to explain the existing phenomenology \citep{bertone}. The possible interplay between ordinary and DM could reveal interesting novel features, thus serving as a smoking gun evidence for the existence of a dark sector. As examples of the previous, one could cite a  possible contribution to the reionization of the Universe and the increase of gas temperature prior to the reionization epoch, leaving a potentially detectable imprint on the cosmological 21-cm signal, as studied in \citet{re1} and \citet{re2}. More in particular, the  production of SM neutrinos from annihilation of proposed dark candidates, generically $\chi$, with energy $E_\nu \lesssim m_\chi$ is of paramount importance for the description of internal dynamics and energetic balance in stellar scenarios. In order to be specific, for example, one can consider the solar context. A DM particle will be gravitationally captured by the Sun if, in scattering against solar nuclei, it falls below the local escape velocity. This accumulation mechanism can lead to a local stellar DM density higher than that of the galactic halo where it resides, potentially providing us an opportune region in which to search for visible signatures \citep{kouv2,aldo, rott}. Additionally, low density environments in solar-type stars could yield interesting features in the neutrino channel \citep{ani,palomares}. 

On the other hand, in denser stellar environments such as those leading to the formation of neutron stars (NSs), neutrinos are vastly produced as they are very efficient at releasing the excess of gravitational energy when a compact stellar object is formed from a more massive progenitor. It has been now more than 30 years since the supernova SN1987A event allowed us to glimpse the complex behaviour of the neutrino internal dynamics and obtain confirmation of the existence of a preliminary neutrino trapping phase followed by a transparency era \citep{gne} from the neutrino telescopes on Earth \citep{yuksel}. In addition, X-ray satellite measurements have also provided indications of the cooling sequence, for a catalog of isolated cooling NSs, see \citep{vigano} and \citep{yako1}. Although a global understanding of the extracted temperatures for these objects is still missing, the so-called minimal cooling mechanism has been successful at reproducing the trends of observed cooling curves \citep{page}. When solving for the internal temperature profile $T(r, t)$ as a function of stellar radius $r$ and time $t$, one of the key ingredients that can dictate the energetic balance is the local energy emissivity, $Q_E=\frac{dE}{dVdt}$, i.e. the energy produced per unit volume per unit time, through a prescribed particle physics reaction.

 In this work we will be interested in obtaining astrophysical neutrino emissivities related to novel reaction channels involving DM undergoing self-annihilation processes inside the star. In particular we focus on models in which dark matter particles communicate with the visible sector through a pseudoscalar mediator. These have been quoted to be well-motivated both from theoretical and from  phenomenological grounds. Some of these models belong to a set of the so-called {\it simplified} type including 
 ~\citet{Boehm:2014hva}, \citet{Wild:2016dds}, \citet{Bauer:2017ota} and \citet{Baek2017vzd}. As mentioned, they extend the SM by (at least) two particles, a DM candidate as well as a state that 
mediates the DM interactions with the visible sector, and are able to capture, with a minimal set of assumptions, some important features of more ultraviolet-complete (UV) theories while providing a (semi-)consistent framework in order to analyse the experimental results \citep{baner}.

In this setting we will be interested in the dominant neutrino production processes, i.e. the s-wave process $\chi \chi \rightarrow \nu \bar{\nu}$ and the p-wave process $\chi \chi \rightarrow a a$, with subsequent decay $a \rightarrow \nu \bar{\nu}$.  Although the previous reactions constitute the main neutrino emission channels in this model setting, additional reactions like e.g. radiative $a$-emission or $\chi \chi \rightarrow a a a$ could also happen but we will not consider them here as they are subdominant. As we will show, the two main reactions could provide a contribution to the standard astrophysical neutrino emissivities in NS environments of sizable magnitude at early times. The observability of such -indirect- effects caused by DM seems nowadays difficult as it could be critically relying on the finest capabilities of current and future X-ray and gamma satellites (NICER, eXTP, LOFT, ATHENA, CHANDRA).

%Previous works evaluated in a the Let us mention at this point that some works predict a warming effect in central regions of the galactic halo for DM densities $\sim 10^3$ \citet{kouv2}.
%%%%%%%%%%%%%
%%%%%%%%%%%%%
%%% SECTION 2 %%%
%%%%%%%%%%%%%
%%%%%%%%%%%%%
\bigskip

\section{Neutrino emissivities from DM annihilation}
\bigskip
%%%%%%%%%%
 In this work we are interested in calculating neutrino emissivities from DM self-annihilation in dense and hot stellar interiors, i.e. that of a (proto) NS. In order to carry out our calculation we choose a model where DM particles interact with SM particles  through a pseudoscalar mediator. This kind of models are well-motivated, both from the  theoretical and phenomenological point of view. With direct detection bounds being typically subleading in such scenarios, the main constraints arise from collider searches (meson bounds) and  from indirect detection experiments \citep{baner}. Examples recently used along this line include the coy dark matter model of ~\citet{Boehm:2014hva} and others \citep{Wild:2016dds, Bauer:2017ota, Baek2017vzd}. Although popular, we must stress that these simplified models have some limitations, regarding construction itself, and  when confronted with bounds from collider searches \citep{gauge, meson}. 

We now introduce our concrete model realization. We consider a model where the SM field content is extended by a Dirac fermion, $\chi$, with mass $m_{\chi}$, which plays the role of a dark matter candidate, and a pseudoscalar field, $a$, with mass $m_a$,  which mediates the interaction of ordinary and dark sectors. The interaction lagrangian of the model reads
\begin{equation}
\mathcal{L_I} =  - i \frac{g_{\chi}}{\sqrt{2}} a \bar{\chi} \gamma_5 \chi - i g_0 \frac{g_{f}}{\sqrt{2}} a \bar{f} \gamma_5 f \, ,
\label{li}
 \end{equation}
where $g_{\chi}$ is the DM-mediator coupling, $g_{f}$ corresponds to the couplings to the SM fermions, $f$, and $g_0$ is an overall scaling factor. From the  usual schemes used for matter couplings when introducing Beyond-Standard-Model motivated physics we will restrict for simplicity to the so-called {flavour-universal}, which sets $g_f = 1$ for all SM fermions. Let us recall, however, that there are other schemes where $a$ couples either to quarks or leptons exclusively, and with a flavour structure which will be treated elsewhere.

Typically, in these models DM phenomenology is controlled by four parameters, $m_\chi$, $m_a$, $g_\chi$, and $g_0 g_f$. In the range $m_\chi < m_{\rm Higgs}$ and $m_a < m_\chi$, the relevant annihilation processes  into two-body final states~\citep{abdu, Arina} are s-wave $\chi {\chi} \rightarrow f\bar{f}$ and p-wave $\chi {\chi} \rightarrow a a$. As a remark it is worth mentioning that, as presented, the most straightforward UV-completion of this setup would be in the framework of the two Higgs doublet model or models involving even more extended scalar sectors. However, one should keep in mind that additional interactions with extra scalars  arise at tree-level and that can introduce important phenomenological model-dependent features \citep{haber}.

Despite the limitations of simplified models, in our particular realization it is reasonable to expect that the very light mediators will not distort the relic density predictions due to the presence of additional annihilation channels involving these extra scalars as discussed in \citet{baner}. 

DM abundance in our universe is likely to be fixed by the thermal freeze-out phenomenon: DM particles, initially
present in our universe in thermal equilibrium abundance, annihilate with one another until chemical
equilibrium is lost due to the expansion of the universe. The present-day relic density of these particles is predictable and it has been measured by Planck \citep{relic} to be $\Omega_{\rm CDM}h^2 = 0.1198\pm0.0015$.

Due to the pseudoscalar portal considered here, this model provides spin dependent interactions with nucleons (N) at tree level. In this way the $\chi$-N interaction considered in direct searches is suppressed because it is momentum dependent, see \citet{Freytsis:2010ne}, \citet{Cheng:2012qr} and  \citet{Gresham:2014vja} for details. Instead, the spin independent cross section is not present at tree level but the effective interaction at one-loop can be constructed~\citep{Ipek:2014gua}. Estimations of both cross sections in vacuum are given in~\citet{Freytsis:2010ne}. Both features regarding the behaviour of the cross section impact the capability of the star to capture DM during the stage of progenitor and in the collapsed configuration, although they can compensate each other in the star lifetime in order to have a finite meaningful amount of DM populating the object \citep{kouv3}.

Usual model analysis considers sets of parameters with a variety of bounds at different level of significance. Here, in order to be definite, we will restrict our analysis to three different sets of  flavour-universal parameters that are not in conflict with existing phenomenology to describe light DM ($m_\chi\lesssim 30$ GeV) interactions with ordinary matter. We consider constraints from direct detection experiments \citep{review}, cosmological bounds \citep{cosmo} and collider bounds \citep{meson}. The masses and couplings used in this work appear in Table 1.

%%%%%%%%%%%%%%%%%%%%%%%%%%%%%%%%%%%%%%%%%%%%%%%%%%%%%%%
\begin{deluxetable*}{ccCrlc}[t!]
\tablecaption{Parameters used in this work as appearing in the interaction lagrangian in Eq. (\ref{li}). \label{tab:mathmode}}
\tablecolumns{6}
\tablenum{1}
\tablewidth{0pt}
\tablehead{
\colhead{Model \tablenotemark{a} } &
\colhead{$m_\chi$ [GeV]} & 
\colhead{ $m_a$ [GeV]} & \colhead{ $g_{\chi}$} & \colhead{$g_0$} \\
}
\startdata
 A & 0.1 & 0.05 & $7.5 \times 10^{-3}$ & $7.5 \times 10^{-3}$ \\
B & 1 & 0.05 & $1.2 \times 10^{-1}$ & $2 \times 10^{-3}$ \\ 
C & 30 & 1 & $6 \times 10^{-1}$ & $5 \times 10^{-5}$ \\
\enddata
\tablenotetext{a}{ We use flavour-universal  $g_f=1$.}
\end{deluxetable*}
%%%%%%%%%%%%%%%%%%%%%%%%%%%%%%%%%%%%%%%%%%55
Model sets A and B are mainly determined by DM relic abundance since the dark candidate mass is in the region where direct detection experiments are less restrictive \citep{Ipek:2014gua}. The couplings in set C are chiefly constrained by LUX results \citep{lux} in spin independent and spin dependent cross sections and, in addition, they respect restrictive rare meson decays \citep{meson} as well. In the beforementioned cases we estimate the parameters using  MicroOmegas \citep{micro} and direct detection cross sections at one-loop level \citep{Freytsis:2010ne,Ipek:2014gua}.

According to the current stage of exploration of the phase space of masses and cross sections for DM candidates interacting  with nucleons in ordinary matter, dense compact stars are believed to be suitable places to find this kind of matter. NSs are believed to be efficient DM accretors \citep{gould}. One of the key quantities that can dictate their internal stellar energetic balance is the local energy emissivity, $Q_E=\frac{dE}{dVdt}$ (energy produced per unit volume per unit time, through a prescribed particle physics reaction). In this work we will be interested in the annihilation reactions of DM into two-body fermionic states ($f$), $\chi {\chi} \rightarrow f\bar{f}$ and two pseudoscalar boson states $\chi {\chi} \rightarrow a a$ with subsequent decay $a \rightarrow f\bar{f}$. Furthermore, we will discuss possible astrophysical consequences particularizing to the $f=\nu$ neutrino channel. 

\begin{figure}[t]
\centering
\includegraphics[scale=1.5,width=0.3\columnwidth]{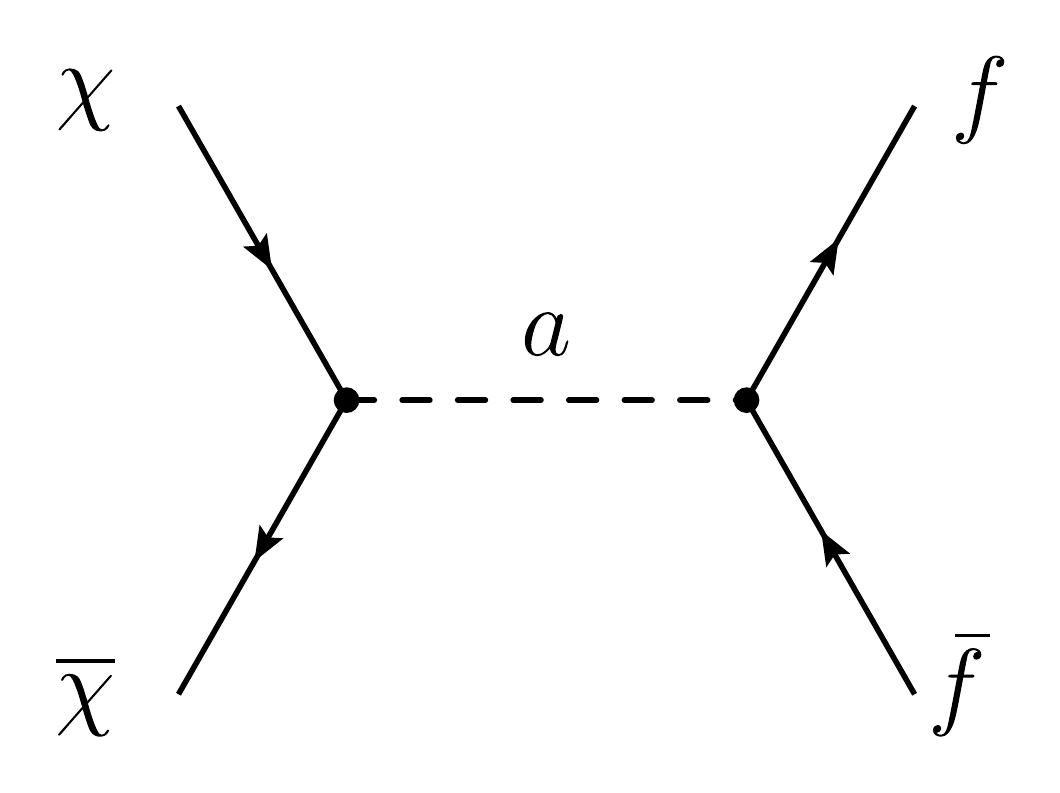}
\includegraphics[scale=1.5,width=0.65\columnwidth]{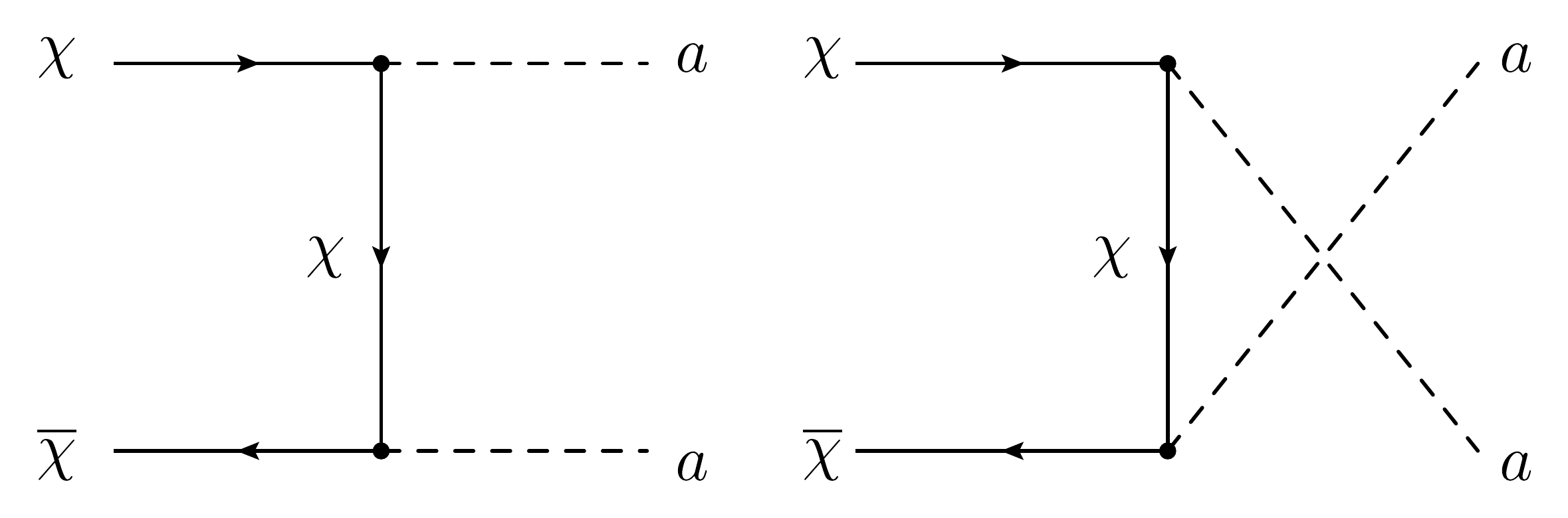}
\caption{Feynman diagrams for DM annihilation  reactions $\chi {\chi} \rightarrow f\bar{f}$ and $\chi {\chi} \rightarrow a a$ considered in this work.}
\label{Feynman2}
\end{figure}

Formally, the expression for $Q_E$ generically denotes the energy emission rate  per stellar volume arising from fermionic or pseudoscalar pair production and can be written as \citep{Esposito}
\begin{equation}
Q_E= 4\int d \Phi (E_1+E_2)\, |\overline{\mathcal{M}}|^2\, f(f_1,f_2,f_3,f_4),
\label{qe}
\end{equation}
with
\begin{equation}
d\Phi=\frac{d^3\vec{p_1}}{2(2 \pi)^{3}E_1} \frac{d^3\vec{p_2}}{2(2 \pi)^{3}E_2} 
\frac{d^3\vec{p_3}}{2(2 \pi)^{3}E_3} \frac{d^3\vec{p_4}}{2(2 \pi)^{3}E_4}\, (2 \pi)^4 \delta (p_1+p_2-p_3-p_4),
\end{equation}
 the 4-body ($12\rightarrow 34$) phase space element and $|\mathcal{\overline{M}}|^2 $, the spin-averaged squared matrix element of the reaction considered. The additional factor $f(f_1,f_2,f_3,f_4)$ accounts for the global phase space blocking factor due to the initial and final particle distribution functions, $f_i, \,i=1,...,4$ we will discuss below. $\delta(x)$ is the 4-dimensional delta function. We will denote $p_1=(E_1, \vec{p_1})$, $p_2=(E_2, \vec{p_2})$ as the incoming 4-momenta, while $p_3=(E_3, \vec{p_3})$, $p_4=(E_4, \vec{p_4})$ are the outgoing 4-momenta, respectively. The detailed associated Feynman diagrams are shown in Fig.(\ref{Feynman2}). Let us note that besides the quoted annihilation processes we consider, there may be additional pseudoscalar boson production s-wave $\chi {\chi} \rightarrow aaa$ \citep{abdu}, initial/final state radiation and internal bremsstrahlung processes $\chi {\chi} \rightarrow f\bar{f}a$ or  $\chi {\chi}a \rightarrow f\bar{f}$ \citep{bell}. 
 However, since the cross sections for these processes are proportional to $g_\chi^2 g_f^4$ and  $g_\chi^4 g_f^2$, respectively, they are subdominant in the case of a Dirac fermion DM candidate \citep{bring1, ibarra1}. Similarly, radiative $a$-production can arise from the SM particles interaction inside the star, but this process is found to be only relevant in the case of very light mediators ($\lesssim$~eV) like axions or Majorons \citep{seda,farzan}.

%%
%%%%%%%%%%%%%%%%%%%%%%%%%%%%%%%%%%%%%%%%%%%%%%%%%%%
%%%%%%%%%%%%%%%%%%%%%%%%%%

Specifically, for the case of annihilation into fermionic pairs (left diagram in Fig.(\ref{Feynman2})) we label the emissivity as $Q_E^{f\bar{f}}$. It  includes the expression for the spin-averaged squared matrix element as 
\begin{equation}
|\mathcal{\overline{M}}_{f\bar{f}}|^2=\frac{g_\chi^2g_f^2}{4}\frac{s^2}{(s-m_a^2)^2+E_{|\vec{q}|}^2\Gamma^2},
\label{M}
\end{equation}
where $q^2=s=(p_1+p_2)^2=(p_3+p_4)^2$ is the Mandelstam variable and $E_{|\vec{q}|}=\sqrt{|\vec{q}|^2+m_a^2}$. In this case 
\begin{equation}
f(f_1,f_2,f_3,f_4) =f_{\chi}(E_1)f_{\bar{\chi}}(E_2)(1-f_{f}(E_3))(1-f_{\bar{f}} (E_4)),
\end{equation}
and $f_\chi,f_{f}$ are the local stellar distribution functions for DM and fermionic particles, respectively, containing density and temperature dependence we will discuss further below.
$\Gamma$ is the pseudoscalar particle decay width in the local medium through the reaction $a \rightarrow f\bar{f}$. It is obtained using the optical theorem  as 
\begin{equation}
\Gamma=\frac{1}{E_{|\vec{q}|}}{\rm Im}\Pi(\vec{q}),
\end{equation}
where $\Pi(\vec{q})$ is the pseudoscalar polarization insertion  given by 
\begin{equation}
\Pi(\vec{q})=\frac{i g_f^2}{2}\int \frac{d^4k}{(2\pi)^4}tr[\gamma_5G^0(k)\gamma_5G^0(k+q)],
\label{pi}
\end{equation}
and the corresponding cut of the associated tadpole diagram involves the fermion propagator $G^0(k)$ including a vacuum and matter contribution \citep{Chin, Matsui}.
%\begin{equation}
%G^0(k)=(\slashed{k}+m_f)\left[\frac{1}{k^2-m_f^2+i\epsilon}+ 2\pi i\delta(k^2-m_f^2) f_f(k^0) %\Theta(k^0)\right]
%\end{equation}
% being $\Theta (k^0)$ the Heaviside function.

Using the Eq.(\ref{qe}) and Eq.(\ref{M}) we can obtain an expresion for the emissivity produced by the annihilation of DM particles into ${f\bar{f}}$, $Q_E^{f\bar{f}}$. Let us first deal with the integration over $\vec{p_4}$ so that
\begin{equation}
\int \frac{d^3\vec{p_3}}{2E_3(2\pi)^3}\int \frac{d^3\vec{p_4}}{2E_4(2\pi)^3}(2\pi)^4\delta (p_1+p_2-p_3-p_4)= \int \frac{2\pi |\vec{p_4}|^2d|\vec{p_4}|}{4E_3E_4(2\pi)^2}d(\cos\,\theta)\delta(q_0-E_3-E_4)
\end{equation}
where $\theta$ is the angle between $\vec{p_4}$ and $\vec{q}$, and $q_0=E_1+E_2$ for this annihilation channel.
Besides, we can express the energy delta function as 
\begin{equation}
\delta(q_0-E_3-E_4)=\frac{\sqrt{m_f^2+|\vec{q}|^2+|\vec{p_4}|^2-2|\vec{p_4}||\vec{q}|\cos\, \theta}}{|\vec{p_4}||\vec{q}|} \delta(\cos\, \theta -\cos\, \theta_{0}) \Theta(q^2\geq 4m_f^2),
\label{eqdelta}
\end{equation}
where
\begin{equation}
\cos\; \theta_{0}=\frac{1}{2|\vec{p_4}||\vec{q}|}\left( |\vec{q}|^2-q_0^2+2q_0E_4\right) ,
\label{cos}
\end{equation}
and $q^2=q_0^2-|\vec{q}|^2$. $\Theta(x)$ is the Heaviside function.
%\begin{equation}
%Q=\frac{1}{16(2\pi)^7}\int \frac{d^3\vec{p}}{2E}\int\frac{d^3\vec{p'}}{2E'}\int_{|\vec{p}|-|\vec{p'}|}^{|\vec{p}|+|\vec{p'}|}d|\vec{q}|\int_{2m_f}^{\infty}d\omega\delta(E+E'-\omega-\omega')\frac{(E+E')|\mathcal{\overline{M}}|^2|\vec{q}|}{|\vec{k'}|\omega'}\mathcal{F}S(q_0)
%\label{eqQ}
%\end{equation}
%where $\mathcal{F}=f^-(E)f^+(E')(1-f^-_f(\omega))(1-f^+_f(\omega'))$.\\
Eq.(\ref{eqdelta}) has been obtained using
\begin{equation}
\delta[f(x)]=\sum_{i} \frac{\delta(x-x_{0i})}{|f'(x)|_{x_0i}|},
\label{delta}
\end{equation}
with  $x_{0i}$ the zeros of $f(x)$. Now, we use that $|\vec{p_4}|d|\vec{p_4}|=E_4dE_4$. Imposing $\cos^2\theta_0\leq 1$, we obtain limits for the integration over $E_4$
\begin{equation}
E_4\pm=\frac{1}{2}\left(q_0 \pm |\vec{q}|\sqrt{1-\frac{4m_f^2}{q^2}}\right).
\end{equation}
In the same way, we use $|\vec{p_1}|d|\vec{p_1}|=E_1dE_1$ and $|\vec{p_2}|d|\vec{p_2}|=E_2dE_2$. After that, Eq.(\ref{qe}) takes the form
%\begin{equation}
%Q_{f\bar{f}}=\frac{1}{2(2\pi)^5} \int_{m_\chi}^\infty dE_1 \sqrt{E_1^2-m_{\chi}^2}\int_{m_\chi}^\infty dE_2 \sqrt{E_2^2-m_{\chi}^2}\int_{-1}^{1}d(cos\phi)q_0\Theta(q^2\geq 4m_f^2)\int_{E_4-}^{E_4+}dE_4 \int_{-1}^1 dcos\theta \delta(cos\theta-cos\theta_0)\mathcal{F}\frac{|\mathcal{\overline{M}}_{f\bar{f}}|^2}{|\vec{q}|},
%\end{equation}
\begin{eqnarray}
Q_E^{f\bar{f}} \nonumber & = & \frac{1}{2(2\pi)^5} \int_{m_\chi}^\infty dE_1 \sqrt{E_1^2-m_{\chi}^2}\int_{m_\chi}^\infty dE_2 \sqrt{E_2^2-m_{\chi}^2}\int_{-1}^{1}d(\cos\,\phi)q_0\Theta(q^2\geq 4m_f^2)\\ 
 & & \times  \int_{E_4-}^{E_4+}dE_4 \int_{-1}^1 d\cos\theta \delta(\cos\,\theta-\cos\,\theta_0)f(f_1,f_2,f_3,f_4)\frac{|\mathcal{\overline{M}}_{f\bar{f}}|^2}{|\vec{q}|}
\label{Qff}
\end{eqnarray}
where $\phi$ is the angle between $\vec{p_1}$ and $\vec{p_2}$, 
$
E_3=E_1+E_2-E_4,
$
\begin{equation}
|\vec{q}|=\sqrt{|\vec{p_1}|^2+|\vec{p_2}|^2+2|\vec{p_1}||\vec{p_2}|\cos\,\phi},
\end{equation} 
and $|\vec{p_1}|=\sqrt{E_1^2-m_\chi^2}$, $|\vec{p_2}|=\sqrt{E_2^2-m_\chi^2}$.

Instead, for the DM annihilation into two pseudoscalars (middle and right diagrams in Fig.(\ref{Feynman2})) now the emissivity is labeled as $Q_E^{a{a}}$. In this case, when calculating the spin averaged matrix element one must note that 
\begin{equation}
|\mathcal{\overline{M}}_{aa}|^2=\frac{1}{2}\frac{1}{2}\sum_{s,s'}|\mathcal{M}_{aa}|^2,
\end{equation}
where $s$ and $s'$ are the spin states of the dark matter particle. The squared matrix element finally reads,
\begin{eqnarray}
|\mathcal{\overline{M}}_{aa}|^2&=&\frac{-g_\chi^4}{2} \lbrace \frac{(t-m_a^2)^2-m_\chi^2(m_\chi^2+2m_a^2)}{(t-m_\chi^2)^2}  + \frac{(u-m_a^2)^2-m_\chi^2(m_\chi^2+2m_a^2)}{(u-m_\chi^2)^2} +2\frac{2m_\chi^2-s}{u-m_\chi^2} \nonumber \\ & + &  \frac{(s-2m_\chi^2)(2m_a^2-s)+2m_\chi^2(m_\chi^2+2m_a^2-2s)-2(t-m_a^2)^2}{(t-m_\chi^2)(u-m_\chi^2)} \rbrace,
\end{eqnarray}
where  $s=k^2=(p_1+p_2)^2$, $t=(p_1-p_3)^2=(p_4-p_2)^2$ and  $u=2m_\chi^2+2m_a^2-s-t$. 

As we can see, now the matrix element not only depends on $s$ but also on $t$ and $u$. Dealing with the integration to obtain the emissivity it is convenient to write these variables as
\begin{equation}
s=2m_\chi^2+2E_1E_2-2|\vec{p_1}||\vec{p_2}|\cos\,\theta_{12},
\end{equation}
and
\begin{equation}
t=m_\chi^2+m_a^2-2E_1E_3+2|\vec{p_1}||\vec{p_3}|\cos\,\theta_{13},
\end{equation}
being $\theta_{ij}$ the angle between $\vec{p_i}$ and $\vec{p_j}$.

In the same way that we did for the annihilation into fermions we can write
\begin{equation}
\int \frac{d^3\vec{p_3}}{2E_3(2\pi)^3}\int \frac{d^3\vec{p_4}}{2E_4(2\pi)^3}(2\pi)^4\delta (p_1+p_2-p_3-p_4)= \int \int \int \frac{|\vec{p_3}|^2d|\vec{p_3}| d\phi_3}{4E_3E_4(2\pi)^2}d(\cos\,\theta_3)\delta(k_0-E_3-E_4),
\end{equation}
where we are denoting the four momentum $k=(k_0, \vec{k})$ and $\theta_i$ as the angle between $\vec{p_i}$ and $\vec{k}$. As obtained in Eq.(\ref{eqdelta}), we find
\begin{equation}
\delta(k_0-E_3-E_4)=\frac{E_4}{|\vec{p_3}||\vec{k}|}\delta(\cos\,\theta_3-\cos\,\theta_{3,0}),
\end{equation}
being
\begin{equation}
\cos\,\theta_{3,0}=\frac{1}{2|\vec{p_3}||\vec{k}|}(|\vec{k}|^2+2k_0E_3-k_0^{2}).
\end{equation}
Now, we can write the emissivity into two pseudoscalars as
\begin{eqnarray}
Q_E^{a{a}}\nonumber & = & \frac{1}{2(2\pi)^6}\int_0^\infty |\vec{k}|d|\vec{k}|\int_{m_\chi}^\infty \sqrt{E_2^2-m_\chi^2} dE_2 \int_{-1}^1 d\cos\,\theta_2 \frac{E_1+E_2}{E_1} \Theta(k^2-4m_a^2)\\ 
 & & \times  \int_{E_3-}^{E_3+} dE_3 \int_{-1}^1 d\cos\,\theta_3 \delta(\cos\,\theta_3-\cos\,\theta_{3,0}) \int_0^{2\pi} d\phi_3 f(f_1,f_2,f_3,f_4) |\mathcal{\overline{M}}_{aa}|^2
\label{Qaa}
\end{eqnarray}
where $|\vec{p_1}|=\sqrt{|\vec{k}|^2+|\vec{p_2}|^2-2|\vec{k}||\vec{p_2}|\cos\,\theta_2}$, 
$t=m_\chi^2+m_a^2-2E_1E_3+2|\vec{p_3}|(|\vec{k}|\cos\,\theta_3-|\vec{p_2}|\cos\,\theta_{23})$ 
and $\cos\,\theta_{23}=\cos\,\phi_3 \,\sin\,\theta_2\,\sin\,\theta_3+\cos\,\theta_2\cos\,\theta_3$. We have also used the trigonometric relation $|\vec{p_1}|\cos\,\theta_{13}=|\vec{k}|\cos\,\theta_3-|\vec{p_2}|\cos\,\theta_{23}$. The limits for the outgoing energy in the integral are
\begin{equation}
E_3 \pm=\frac{1}{2}\left(k_0 \pm |\vec{k}|\sqrt{1-\frac{4m_a^2}{k^2}} \right).
\end{equation}

In the case of annihilation into pseudoscalars the phase space factor reads
\begin{equation}
f(f_1,f_2,f_3,f_4)=f_\chi(E_1)f_{\bar{\chi}}(E_2)f_a (E_3)f_a(E_4).
\end{equation}
In this case one should also take into account the further decay of each pseudoscalar into fermionic pairs and the availability of kinematical phase space through and additional Pauli blocking factor. Although not explicit, there is also a further local dependence on the DM density in the distribution function that will be discussed later in the manuscript.
%0$ as the final state bosons rapidly decay into fermions. Particularizing  (neutrinos) thus we do not expect to have a net stellar neutrino density. 

%%%%%%%%%%%%%
%%%%%%%%%%%%%
%%% SECTION 3 %%%
%%%%%%%%%%%%%
%%%%%%%%%%%%%
\bigskip
\subsection{Dense and hot stellar scenario}
\bigskip
In order to explain the physical relevance of the quantities under scrutiny obtained in the previous section, at this point we will particularize to that of a dense and hot stellar scenario. We will focus on a (proto) NS. Briefly, a NS is mostly constituted by nucleons forming a central core at a density in excess of nuclear saturation density, $\rho_0\simeq 2.4 \times10^{14}\, \rm g/cm^3$. An average NS has a radius $R\lesssim 12$ km and mass $M\sim (1-1.5) M_{\odot}$ (mostly in its core) being thus a star with large compactness ratio $\sim M/R$. For the sake of our discussion we will consider a typical baryonic core density value  $\rho_b=\rho_N\sim 2 \rho_0$. Regarding internal temperature and composition, NSs are born as hot lepton-rich objects with temperatures $T\sim 20$ MeV evolving into cold $T\sim 10$ keV neutron-rich ones, after a deleptonization era. Assuming dark and ordinary matter have coupling strengths at the level of current experimental search bounds, NSs are believed to be capable of accreting (and retaining) DM particles whose masses are larger than a few GeV from an existing galactic distribution.

Accretion of a dark component will proceed not only during the collapsed stage but also during most of the previous progenitor stellar lifetime at different epoch-dependent capture rates, $C_{\chi}$. First, in the progenitor stages, the progressively denser nuclear ash central core is effectively opaque to DM and allows building up an internal finite DM number density over time, $n_\chi(r)$, being $r$ the radial stellar coordinate. Briefly, the progenitor with a mass $\sim (10-15)M_{\odot}$ is able to fuse lighter elements into heavier ones and thus its composition changes through the burning ages. Hydrogen first, and later the He, C, O, and rest of heavier elements up to Si proceed through the burning stages. Spin-dependent (mostly from H) as well as spin independent $\chi$-N cross sections allow  the gravitational capture of  DM population inside the star. Coherence effects may play a role for slowly moving, low ${m_{\chi}}$ incoming DM particles scattering nuclei off when their associated de Broglie wavelength is comparable to the nuclear size, and in this case the spin-independent cross section bears a multiplicative factor $\simeq A^2$ where $A$ is the baryonic number. Since the later burning stages proceed rapidly, the ${\rm He-C-O}$ stage gives the main contribution to the DM capture in the progenitor. As the thermalization times during this set of stages can follow the internal dynamics the collapsed star will have as a result a non-zero, mostly inherited, initial DM  population.

Most in detail, the DM particle population number inside the star, $N_\chi$, will  not only depend on the capture rate $C_\chi$  \citep{gould}  but also on the  self-annihilation rate, $C_a$. Note that in the range of masses in the parameter sets we consider, evaporation effects \citep{evap} as well as decay \citep{decay} do not substantially modify the DM population as the kinetic to gravitational potential energy ratio remains small. 

Then the DM particle number, $N_{\chi}$, can be obtained a function of time $t$ by solving the differential equation
\begin{equation}
\frac{dN_{\chi}}{dt}=C_{\chi}-C_a N^2_{\chi},
\end{equation}
considering the two competing processes, capture and annihilation \citep{kouv1}
\begin{equation}
N_\chi(t)=\sqrt{\frac{C_\chi}{C_a}} \rm \,tanh \left[\frac{t}{\tau}+\gamma (N_{\chi,0})\right],
\label{Nx}
\end{equation}
 where 
\begin{equation}
\gamma (N_{\chi,0})={\rm tanh}^{-1} \left(\sqrt{\frac{C_\chi}{C_a}}N_{\chi,0}\right) 
\end{equation}
 and 
\begin{equation}
\tau^{-1}=\sqrt{C_\chi C_a}.
\end{equation}

At $t=0$, when the protoNS is born, a typical progenitor may have already provided an initial population 
\begin{equation}
 N_{\chi,0} = 1.5\times 10^{39} \left( \frac{\rho_\chi}{\rho^{ambient}_{\chi,0}}\right) \left( \frac{1 \; \rm GeV}{m_\chi}\right) \left(  \frac{\sigma_s}{10^{-43} \; \rm cm^2} \right)  ,
\label{nxo}
\end{equation}
where $\sigma_s\equiv \sigma_{\chi-N}$ is the $\chi-N$ scattering cross section. As this quantity is currently unknown, only experimental constrains exist for it. In the range of DM masses used in this work $\sigma_s\in [10^{-46}-10^{-33}]\,\rm cm^{-2}$ \citep{limits}. Eq. (\ref{nxo}) assumes that the majority of the NS population can be found at galactocentric distances of a few kpc where $\rho_{\chi}\sim 10^2 \rho^{ambient}_{\chi,0}$. We use $\rho^{ambient}_{\chi,0}\simeq 0.3$ $\rm \frac{GeV}{cm^3}$ as the solar-circle DM density value.

Let us mention that both capture and annihilation rates, will be intimately determined by the parameters of the model at hand, i.e. $m_\chi, m_a, g_0, g_\chi$ (we set $g_f=1$). In particular, the DM capture rate on the progenitor depends on the scattering cross section on nuclei (nucleons) that is
proportional to the product of the couplings $(\sim g^2_0 g^2_\chi)$ and the annihilation cross section proportional to the sum of  $(\sim g^2_0 g^2_\chi)$ and $(\sim g^4_\chi)$ terms for the two reactions considered $\chi {\chi} \rightarrow f\bar{f}$ and $\chi {\chi} \rightarrow a a$, respectively \citep{buck}.

 For the three models considered in this work appearing in Table 1, the average progenitor capture rate allows a non-vanishing  initial DM population, $N_{\chi,0}$, since the annihilation rate, proportional to $n^2 _\chi(r),$ is negligibly small at that stage. Later, in the NS collapsed state and at a given galactic location with a corresponding ambient DM density $\rho_\chi$, the capture rate, $C_\chi$, it is approximated up to factors of order unity by the expression \citep{gould,lavallaz}.
\begin{equation}
C_{\chi}\simeq 1.8 \times 10^{25} \left(\frac{1\, \rm GeV}{m_{\chi}}\right)\left(\frac{\rho_{\chi}}{\rho^{ambient}_{\chi,0}}\right) f_{\chi,N}\,\,\rm s^{-1}.
\end{equation}
\normalsize
 A few remarks are due regarding this expression. $f_{\chi,N}$ denotes a phenomenological factor dealing with the opacity of stellar matter. $ f_{\chi,N}$ depends on the ratio of the leading contribution of $\chi-N$ scattering cross section $\sigma_{s}$ to the minimum geometrical cross section of a NS made of nucleons of mass $m_N$ and  defined as
\begin{equation}
\sigma_0=\frac{m_N R^2}{M}\sim 10^{-45}\rm cm^2.
\end{equation}

Thus this factor saturates to unity, $f_{\chi,N} \sim 1$, if  ${\sigma_{s}}\gtrsim {\sigma_0}$. Otherwise, $f_{\chi,N} \sim \frac{\sigma_{s}}{\sigma_{0}}$. Using \cite{Boehm:2014hva} and appendix D in \cite{meson} we consider the scattering cross section (at one-loop) in the appropriate kinematical limit in our compact star so that for the parameters used in this work $f_\chi$ is effectively in the saturated regime. 

The expressions in the literature for DM capture rates in dense objects are based on interactions with quarks (nucleons) that in practice happen via a contact term \citep{Freytsis:2010ne}, possibly including form factors. Note that a more ellaborate treatment would involve the calculation of the non-relativistic limit of the (full) series of operators included in the lagrangian under study. Such a detailed analysis is beyond the scope of this work and remains to be done. The usual  phenomenological treatment, through the $f_{\chi,N}$ factor, makes use of a lower  bound to the  global cross section with all relevant contributions in this realization picture. It is important to emphasize that the strength of the computed emissivity will depend on the number of dark matter particles remaining inside in the star at any given time.  

As thermalization times for DM particles in the light mass range we consider are consistently smaller than dynamical cooling times \citep{goldman}, inside the star the DM particle number density takes the form 
\begin{equation}
n_{\chi}(r)=n_{0,\,\chi} e^{-\frac{m_{\chi}}{k_B T}\Phi (r)},
\end{equation}
 with $n_{0,\,\chi}$ the central value, $T$ the NS temperature and $k_B$ the Boltzmann constant. The gravitational potential is given by
\begin{equation}
\Phi(r)=\int_0^r \frac{GM(r') dr'}{{r'}^2},
\end{equation} 
where $M(r')$ is the NS mass inside a spherical volume of radius $r'$.
So that assuming an approximately constant density core,
\begin{equation} 
n_{\chi}(r)=n_{0,\,\chi} e^{-(r/r_{\rm th})^2},
\end{equation} 
 with a thermal radius 
\begin{equation} 
r_{\rm th}= \sqrt{\frac{3 k_B T}{2 \pi G \rho_N m_{\chi}}}.
\end{equation} 
 Normalization requires $\int_0^R n_{\chi}(r) dV=N_\chi$ at a given time, as reflected by Eq.(\ref{Nx}). Note that potential limiting values of $N_\chi$ may arise from the fact that a fermionic $\chi$ would involve the existence of a Chandrasekhar critical mass for collapse \citep{ch}. This possibility is safely not fulfilled as long as $N_{\chi}(t)< N_{\rm Ch}$, where $N_{\rm Ch}\sim (\,M_{\rm Pl}/m_{\chi})^3\sim 1.8 \times 10^{57}\,( 1 {\rm \,GeV}/m_{\chi})^3$ with $M_{\rm Pl}$ the Planck mass. 

Let us now comment on the fact that inside the star the tiny DM fraction can be described by a distribution function of a classical Maxwell-Boltzmann type 
\begin{equation} 
f_\chi=f^{MB}_\chi(|\vec{p_i}|,r)=\left( \frac{1}{2\pi m_\chi k_B T}\right)^{\frac{3}{2}}n_\chi (r) e^{\frac{-|\vec{p_i}|^2}{2m_\chi T}}, \,i=1,2.
\end{equation} 

and in the non-relativistic scheme $E_i=\frac{\vec{p_i}^2}{2m_\chi},\, \rm i=1,2$. The annihilation rate, $C_a$, depends on the thermally averaged annihilation cross section inside the star, $\langle{\sigma_a v}\rangle$, for the two reactions considered in this work, see Fig.(1). Therefore, the stellar $\chi$-distribution contained in the thermal volume region $\sim r^3_{\rm th}$ determines the annihilation rate $C_a \sim \langle{\sigma_a v}\rangle/V_{\rm th}$ \citep{gauge, Arina}. 
 Note that the presence of the phase space factor $f(f_1,f_2,f_3,f_4)$ in Eq. (\ref{qe}) will introduce further DM density and $T$ dependence into the vacuum standard calculation as a thermalized DM distribution exists inside the NS core. As for the outgoing fermions, the medium density effects will generally arise from the phase space blocking factors and collective effects \citep{cer1}. In case of neutrinos we assume $f_\nu\sim 0$, although in cases where a trapped fraction $Y_\nu>0$ exists it would further decrease the response.

%Parameter sets shown in Table I fit the required DM phenomenology, however one should bear in mind that the mechanism described in this work is fed by the DM accretion and will proceed at the rate dictated by dynamical capture and annihilation.

%%%%%%%%%%%%%%%%%%%%%%%%%%%%%%%%%%%%%%%
\begin{figure*}[ht]
  \centering
{\includegraphics[scale=2.0]{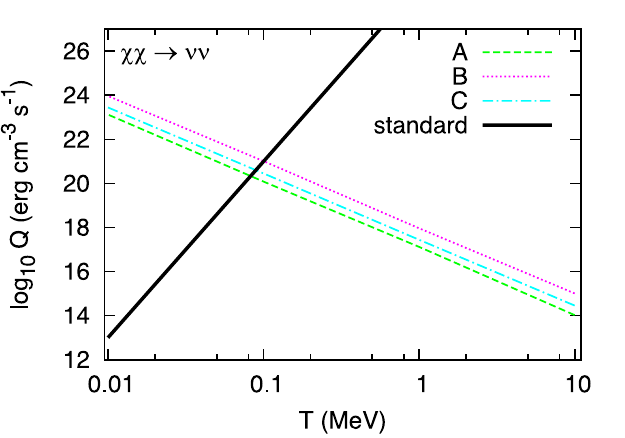}}
\caption{Energy emissivity from DM self-annihilation channel $\chi \chi \rightarrow \nu \nu$ as a function of temperature.  Standard emission refers to MURCA processes. $N_\chi=N_{0,\chi}$ is assumed. See text for details.}
\label{Fig1}
\end{figure*}
%%%%%%%%%%%%%%%%%%%%%%%%%%%%%%%%%%%%%%%

\section{Results}
\bigskip
In this section we explain our results regarding the emissivities in the NS astrophysical scenario particularizing to the case in which  the final state fermions produced in the reactions depicted in Fig.(\ref{Feynman2}) are neutrino pairs. Neutrinos are weakly interacting SM fermions known to play a key role in the internal energy  dynamics of a massive stellar progenitor undergoing gravitational collapse. In such an event most of the gravitational binding energy is emitted into neutrinos (and antineutrinos) of the three families. A very efficient cooling scenario emerges in the first $\sim 10^5$ yr. Standard processes such as those present in the URCA or the modified URCA (MURCA) cooling \citep{friman, yako} among others can release neutrinos with associated emissivities  $Q^{\rm URCA}_E\sim 10^{27} \mathcal{R} (\frac{T}{0.1 \, \rm MeV})^6$ $\rm erg\, cm^{-3}\, s^{-1}$ and $Q^{\rm MURCA}_E\sim 10^{21} \mathcal{R} (\frac{T}{0.1 \, \rm  MeV})^8$ $\rm erg \,cm^{-3} \,s^{-1}$, respectively. Typical energetic scales can be obtained from the conversion factor $1$ MeV $\sim 10^{10}$ K. $\mathcal{R}$ is a reduction function of order unity describing the superfluid effects in the neutron and proton branches of those reactions \citep{yako1}. We must keep in mind the fact that these neutrinos effectively cool off the star as they leave, having scattered a few times with ordinary nucleon matter \citep{horo,ang} after a first rapid trapping stage. In this way processes with neutrino production in reactions involving nucleon components effectively release energy from the baryonic system as the associated neutrino mean free path is relatively long $\lambda\sim 28 \,\rm cm\, (100\,\rm MeV/E_\nu)^2$.  In analogy with what happens at the standard neutrino trapping stage in very young stars, when neutrinos have energies of dozens of MeV, it is expected that energetic neutrinos produced in reactions of DM annihilation could have mean free paths very small, even at low stellar temperatures (in evolved stars) so they may not escape so easily the dense medium \citep{kouv2}.
%%%%%%%%%%%%%%%%%%%%%%%%%%%%%%%%%%%%%%%%%%%%%%%%%%
\begin{figure*}[ht]
  \centering
  \subfigure{\includegraphics[scale=2.0]{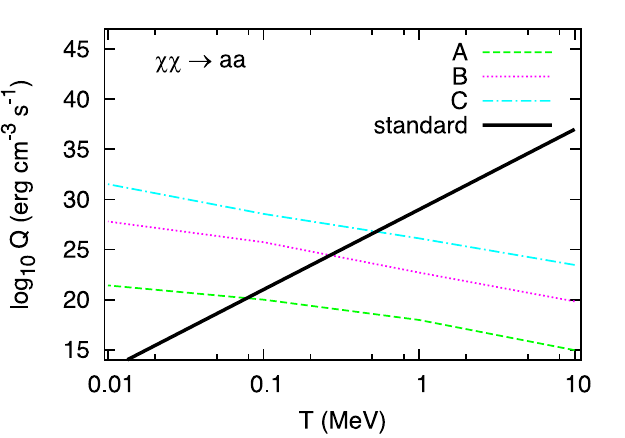}}
\caption{Energy emissivity from DM self-annihilation channel $\chi \chi \rightarrow a a$  with subsequent decay $a \rightarrow \nu \nu$ as a function of temperature. Standard emission refers to MURCA processes. $N_\chi=N_{0,\chi}$ is assumed. See text for details. }
\label{Fig2}
\end{figure*}
%%%%%%%%%%%%%%%%%%%%%%%%%%%%%%%%%%%%%%%

%%%%%%%%%%%%%%%%%%%%%%%%%%%%%%%%%%%%%%%
%\begin{figure*}[t]
%  \centering
%  \subfigure{\includegraphics[scale=1.2]{QE_xxff_T.eps}}\quad
%  \subfigure{\includegraphics[scale=1.2]{QE_xxaa_T.eps}}
%\caption{Energy emissivity from channels $\chi \chi \rightarrow \nu \nu$ (left) and $\chi \chi \rightarrow a a$ (right) with subsequent decay $a \rightarrow \nu \nu$ as a function of temperature. Standard emission refers to MURCA processes.}
%\label{Fig1}
%\end{figure*}
%%%%%%%%%%%%%%%%%%%%%%%%%%%%%%%%%%%%%%%

In Fig.(\ref{Fig1}) the logarithm (base 10) of the energy emissivity for the process $\chi \chi \rightarrow \nu \bar{\nu}$ is shown as a function of temperature for the three sets of DM parameters A, B, C in Table 1, with dashed, dotted and dash-dotted lines, respectively. Baryonic density is fixed at $2\rho_0$. We also fix (for reference) the assumed number of DM particles at initial instant $N_\chi=N_{0,\chi}$. Note, however, that $N_\chi$ is time (and model) dependent as it rapidly decreases when the self-annihilation sets in. We give below a suitable fit where the actual T and $N_\chi$ dependence is reflected. The baryonic density dependence is, however, weak. The standard physics cooling is depicted here by the  MURCA emissivity (solid line) for the sake of comparison. Although the latter is not the only process that could possibly contribute to the effective standard cooling, all other processes capable of are considered weaker at the temperature and density conditions considered in this work. Thus we take the standard processes to be represented by an upper limit to the currently used emissivities chosen as the MURCA processes. Note we do not consider exotic meson codensates nor URCA emissivities either, since stellar central densities required are usually higher than the one taken as reference here in order to provide the $Y_p \sim 11-15\%$  proton fraction \citep{proton} to sustain the fast reaction. Further, we assume that a possible neutrino trapping phase will not be significant, particularly at low T. However, at high T it may produce a further reduction that has to be accounted for through a Pauli blocking factor. It must be included to take into account the time dependent non-vanishing leptonic fraction of order $Y_L\sim 0.1$ \citep{pons} until transparency sets in at $\sim 20$ s.

The trends depicted in Fig.(\ref{Fig1}) with temperature and DM particle population dependence for models A, B and C and for both reaction channels can be fit as

\begin{equation} 
Q_E(T, N_\chi)=Q_0\left( \frac{N_\chi}{N_{0,\chi}}\right)^2 \left(\frac{T}{1\,\rm MeV}\right)^{-3}.
\label{fit}
\end{equation}

In Table 2 we give values for parameters $Q_0$ and $N_{0,\chi}$ for the reactions and models considered. We can see that around $T\sim 0.1$ MeV standard emissivities $\rm log_{10}\, (Q_E)\sim 21$ are as powerful as those from the DM annihilation processes in the thermal volume region $V_{\rm th}\sim r^3_{\rm th}$. However, since DM population is a  decreasing function of time, one can expect that there will be a  minimal number of $N_\chi$ population to beat the MURCA processes for $T <0.1$ MeV. We find $Q_E(T,N_\chi) >Q_{\rm MURCA}$ for $T \in[0.01,0.1]$ $\rm MeV$ for $N_{\chi}/N_{0,\chi} \gtrsim  10^{-5}, 3.6 \times 10^{-6}, 5.6 \times 10^{-6}$ for A, B and C models, respectively. One should note that the $N_\chi$ self-consistently depends on temperature and how it dynamically changes with time. A fully detailed cooling simulation would yield the temporal sequence to determine the complete behaviour. As this is not the goal here, we give instead an estimate on the time duration of the dominance of the DM annihilation channel, i.e. where it could beat the local MURCA processes from Eq.(\ref{Nx}). We obtain $t \lesssim 50$ s for models A and B while this condition is true at all times for model C. We note that all timescales are thus overlapping the  standard transparency window for SM neutrinos. 

In Fig.(\ref{Fig2}) the logarithm (base 10) is shown for the reaction $\chi \chi \rightarrow a a$ with subsequent decay $a\rightarrow \nu \nu$. The number of DM particles is also fixed $N_{\chi}=N_{0,\chi}$ for reference using the same argument as with the $\chi \chi \rightarrow \nu \nu$ reaction in Fig.(\ref{Fig1}). In this case, for a population $N_{0,\chi}$, the neutrino emissivity is largely enhanced with respect to the direct production of neutrinos $\chi \chi \rightarrow \nu \nu$. For model C (dash-dotted line) $Q_E$ matches and surpasses the standard MURCA emission below  $T\sim 0.5$ MeV while for models B (dotted line) and A (dashed line) that happens   for $T\sim 0.3$ MeV and $T\sim 0.1$ MeV, respectively.  Values for the phenomenological fit in this channel are also provided in Table 2. 

If we now consider the running character of DM population number as  in the previous case we find $Q_E(T,N_\chi) >Q_{\rm MURCA}$ in the interval $T \in[0.01,0.1]$ $\rm MeV$ for $N_{\chi}/N_{0,\chi} \gtrsim 3 \times 10^{-5}$ and model A, while for model B $Q(T,N_\chi) >Q_{\rm MURCA}$ in $T \in[0.01,0.3]$ $\rm MeV$ when $N_{\chi}/N_{0,\chi} \gtrsim 5 \times 10^{-8}$. Finally  $Q(T,N_\chi) >Q_{\rm MURCA}$ in $T \in[0.01,0.5]$ $\rm MeV$  for $N_{\chi}/N_{0,\chi} \gtrsim 3 \times 10^{-10}$ for model C. This last ratio for model C is achieved during the entire lifetime of the star while not for the other cases. For the bi-pseudoscalar production reaction and at the thermodynamical conditions in the scenario considered we obtain that typical values obtained for the $a-$decay length are of order of $\sim 10^2$ fm, making it a negligible contribution to the neutrino transport as their decay length is so tiny compared to stellar size.

%%%%%%%%%%%%%%%%%%%%%%%%%%%%%%%%%%%%%%%%%%%%%%%%%%%%%%%
\begin{deluxetable*}{ccCrlc}[t!]
\tablecaption{Parameters obtained for the fit in Eq. (\ref{fit}). \label{tab:mathmode}}
\tablecolumns{6}
\tablenum{2}
\tablewidth{0pt}
\tablehead{
\colhead{Channel} &
\colhead{Model} &
\colhead{ $\rm log_{10} \,Q_0$ $[\rm erg\, cm^{-3}\,s^{-1}]$} & \colhead{$N_{0,\chi}$} \\
}
\startdata
$\chi \chi \rightarrow \nu \nu $ & A & 17.3 &  $4.1 \times 10^{44}$\\ 
$\chi \chi \rightarrow \nu \nu $ & B& 18&  $2.4 \times 10^{44}$ \\
$\chi \chi \rightarrow \nu \nu $ & C & 17.6 &  $5 \times 10^{38}$\\
$\chi \chi \rightarrow a a \tablenotemark{*}$ & A& 18&  $4.1 \times 10^{44}$\\
$\chi \chi \rightarrow a a$ & B & 22.5&  $2.4 \times 10^{44}$ \\
$\chi \chi \rightarrow a a $ & C& 27&  $5 \times 10^{38}$ \\
\enddata
\tablenotetext{*}{ with subsequent decay $a\rightarrow \nu \nu$}
\end{deluxetable*}
%%%%%%%%%%%%%%%%%%%%%%%%%%%%%%%%%%%%%%%%%

 At this point it is worth noting that tighter restrictions in the validity of the coupling of DM to u-d-s quarks coming from including complementary experimental bounds, e.g. rare meson decays, could somewhat reduce the validity of models A and B. In addition, from isospin considerations, the $\chi$ coupling to neutrons and protons in the NS will also affect somewhat the results as the proton-to-neutron ratio inside the NS core is smaller than unity ($\sim 1/9$). As the ratio of coupling fulfills $|g_p/g_n|>1$ when considering flavour universal fermion couplings \citep{meson} we expect that the computed emissivities could be increased by a factor $\sim 10-100$.
 
We have considered the local emissivities of the novel reaction involving self-annihilating matter in the thermal volume. We must emphasize that since there is no uniform distribution of DM inside the star, neutrinos being produced from nuclear reactions in the majority of the core volume will wash out the dark contribution early, when temperature is high enough. However, later, as temperature decreases there is an effective competition of the very efficient dark central engine (located in a few percent of the core volume) and the colder core emission.

The  radial extent of the DM annihilating inner region is correlated to the ratio of the thermal radius to the NS radius. It is defined as $\xi=\sqrt{2} r_{\rm th}/R$ and indicates the radial fraction where DM particles can be found. Since the crust region has a tiny mass we will not consider this refinement here \citep{cer2}.
For the parametrizations A, B, C analyzed in this work this ratio takes values e.g. $\xi \in[0.03,0.42]$ at $T\sim 1$ MeV, $\xi \in[0.007,0.11]$ at $T\sim 0.1$ MeV and  $\xi \in[0.003,0.04]$ at $T\sim 0.01$ MeV. The volume where the dark emitting region resides shrinks as $\sim \sqrt{T}$.

As thoroughly studied, enhanced emissivities in the medium can have an impact on internal temperatures, temporal cooling sequence and (un)gapped matter phases \citep{stei,page-reddy}. In this regard, recent works \citep{cas1,cas2} quote that the rapid cooling of the Cas A may be an indication of the existence of global neutron and proton superfluidity in the core. In addition, current observations of thermal relaxation of NS crusts indicates that even a small stellar volume fraction where fast neutrino emission reactions can take place would provide distinctive features. More in detail, it has been shown \citep{brown} that even if there is a relatively small local volume where distortion from the standard energetic mechanisms is taking place inside the star, a fast neutrino reaction (Direct URCA) in a volume of $\sim 1\%$ could explain the neutrino luminosities in the cooling curve of some particular objects like MXB 1659-29. 

In the case presented in this work the long-term dark engine reaction $\chi \chi \rightarrow a a$ could provide emissivities $Q_E^{aa}\sim 4\times 10^{22}$ $\rm erg\, cm^{-3}\,s^{-1}$ which result higher than that estimated for the $Q_{MX1659-29}\sim 1.7 \times 10^{21}$ $\rm erg\, cm^{-3}\,s^{-1}$for $T\sim 10^{8}$ K. The stellar volume affected in the annihilating DM mechanism is, nevertheless, much smaller for this range of temperatures but still providing the same powerful emission. Besides the process discussed in this work, other ones such as  rotochemical heating \citep{chem} or hot blobs located at different depths in the crust in young NS \citep{heater} have been also treated in the literature adding more sources of energetic variability based on SM matter.

It remains for further work to explore more exhaustively the precise relation between the model parameters $m_\chi, m_a, g_0, g_\chi$ (and $g_f$) and the energetic efficiency of the emission. This will impact the duration of the dominance of such emission from DM annihilation over standard processes and thus its potential observability. We believe that the qualitatively different picture arising from the DM self-annihilation process inside NS may be worthwhile to explore.

\section{ Conclusions}
We have calculated the energy emissivity of self-annihilating dark matter from an existing stellar distribution into final state SM fermions. A pseudoscalar-mediated DM interaction with the ordinary nucleon matter has been used. Later, as an astrophysically relevant case we have particularized to neutrinos as final states, and we have considered in detail those produced from s-wave channels $\chi \chi \rightarrow \nu \nu$ or via pseudoscalar mediators p-wave $\chi \chi \rightarrow a a$, and subsequent decay $a \rightarrow \nu \nu$. In the inner stellar regions the radiation engine can encompass about $\lesssim 7\%$ of the total stellar volume for $T\lesssim 10^{10}$ K and the energy emissivity can be enhanced orders of magnitude compared  to the MURCA standard neutrino processes for parameter sets respecting constraints of direct detection limits, cosmological bounds or even tighter rare meson decay bounds. We have provided a phenomenological fit of emissivities including dependence of temperature and DM particle number. Taking as reference the usual standard temporal sequence of NS cooling behaviour we expect that, for the models analyzed in this work, model C (with $m_\chi=30$ GeV, $m_a=1$ GeV) could  be effectively active during  the whole life of the star.
Although a detailed solution of the full evolution equation is out of the scope of this work it is reasonable to foresee that the contribution of this new dark mechanism to the set of already known  standard  cooling reactions will drive the star into internal dynamical self-adjustment that is likely to emerge with a distinctive temperature sequence whose observability remains to be properly analyzed in future works.

\acknowledgments

We thank useful discussions with J. Silk, J. Edsj\"{o}, H. Grigorian  and C. Albertus. This work has been supported by  PHAROS Cost action, MINECO Consolider-Ingenio { Multidark} CSD2009-00064, Junta de Castilla y Le\'on SA083P17 and FIS2015-65140-P projects. M. Cerme\~no is supported by a fellowship from the University of Salamanca.

%\label{lastpage}

\end{document}